\documentclass{aa}

\usepackage{psfig}

\begin{document}
%
\def\lfir{$L_{\rm FIR}$}
\def\mabs{M$_{\rm abs}$}
\def\etal{et al.}
\def\cbeta{$c_{\rm H\beta}$}
\def\av{A$_{\rm v}$}
\def\flam{$F_{\lambda}$}
\def\ilam{$I_{\lambda}$}
\def\teff{\ifmmode T_{\rm eff} \else $T_{\mathrm{eff}}$\fi}
\def\lg{$\log g$}
\def\feh{$\mathrm{[Fe/H]}$}
\def\mh{$\mathrm{[M/H]}$}
\def\ltsima{$\buildrel<\over\sim$}
\def\lsim{\lower.5ex\hbox{\ltsima}}

\newcommand{\hii}{H~{\sc ii}}
\newcommand{\ha}{\ifmmode {\rm H}\alpha \else H$\alpha$\fi}
\newcommand{\hb}{\ifmmode {\rm H}\beta \else H$\beta$\fi}
\newcommand{\lya}{\ifmmode {\rm Ly}\alpha \else Ly$\alpha$\fi}
\newcommand{\hei}{He~{\sc i}}
\newcommand{\Hei}{He~{\sc i} $\lambda$4471}
\newcommand{\heii}{He~{\sc ii}}
\newcommand{\Heiiuv}{He~{\sc ii} $\lambda$1640}
\newcommand{\Heiiopt}{He~{\sc ii} $\lambda$4686}
\newcommand{\qh}{\ifmmode q({\rm H}) \else $q({\rm H})$\fi}
\newcommand{\qhe}{\ifmmode q({\rm He^0}) \else $q({\rm He^0})$\fi}
\newcommand{\qhep}{\ifmmode q({\rm He^+}) \else $q({\rm He^+})$\fi}
\newcommand{\Qh}{\ifmmode Q({\rm H}) \else $Q({\rm H})$\fi}
\newcommand{\Qhe}{\ifmmode Q({\rm He^0}) \else $Q({\rm He^0})$\fi}
\newcommand{\Qhep}{\ifmmode Q({\rm He^+}) \else $Q({\rm He^+})$\fi}
\newcommand{\Qhtwo}{\ifmmode Q({\rm LW}) \else $Q({\rm LW})$\fi}
\newcommand{\qrathe}{\ifmmode q({\rm He^0})/q({\rm H}) \else $q({\rm He^0})/q({\rm H})$\fi}
\newcommand{\qrathep}{\ifmmode q({\rm He^+})/q({\rm H}) \else $q({\rm He^+})/q({\rm H})$\fi}
\newcommand{\Qrathe}{\ifmmode Q({\rm He^0})/Q({\rm H}) \else $Q({\rm He^0})/Q({\rm H})$\fi}
\newcommand{\Qrathep}{\ifmmode Q({\rm He^+})/Q({\rm H}) \else $Q({\rm He^+})/Q({\rm H})$\fi}
\newcommand{\Qhave}{\ifmmode \overline{Q}({\rm H}) \else $\overline{Q}({\rm H})$\fi}
\newcommand{\Qheave}{\ifmmode \overline{Q}({\rm He^0}) \else $\overline{Q}({\rm He^0})$\fi}
\newcommand{\Qhepave}{\ifmmode \overline{Q}({\rm He^+}) \else $\overline{Q}({\rm He^+})$\fi}
\newcommand{\Qhtwoave}{\ifmmode \overline{Q}({\rm H}_2) \else $\overline{Q}({\rm H}_2)$\fi}
\newcommand{\Qratheave}{\ifmmode \overline{Q}({\rm He^0})/\overline{Q}({\rm H}) \else $\overline{Q}({\rm He^0})/\overline{Q}({\rm H})$\fi}
\newcommand{\Qrathepave}{\ifmmode \overline{Q}({\rm He^+})/\overline{Q}({\rm H}) \else $\overline{Q}({\rm He^+})/\overline{Q}({\rm H})$\fi}

\def\micron{$\mu$m}
\def\kms{km s$^{-1}$}
\def\kmsmpc{km s$^{-1}$ Mpc$^{-1}$}
\def\cmc{cm$^{-3}$}
\def\erg{ergs s$^{-1}$ cm$^{-2}$ \AA$^{-1}$}
\def\ergs{ergs s$^{-1}$}
\def\ergscm{ergs s$^{-1}$ cm$^{-2}$}
\def\msun{\ifmmode M_{\odot} \else M$_{\odot}$\fi}
\def\zsun{\ifmmode Z_{\odot} \else Z$_{\odot}$\fi}
\def\lsun{\ifmmode L_{\odot} \else L$_{\odot}$\fi}

\def\mup{\ifmmode M_{\rm up} \else M$_{\rm up}$\fi}
\def\mlow{\ifmmode M_{\rm low} \else M$_{\rm low}$\fi}
\def\ubvetc{(UBV)$_J$\-(RI)$_C$\- JHKLL$^\prime$M}
\def\basel {{\it B}a{\it S}e{\it L}}
\def\was{CMT$_1$T$_2$}

%
\def\aap{A\&A}
\def\aaps{A\&AS}
\def\aas{A\&AS}
\def\aj{AJ}
\def\apj{ApJ}
\def\apjl{ApJ}
\def\apjs{ApJS}
\def\mnras{MNRAS}
\def\pasp{PASP}
%


\title{The transition from Population III to normal galaxies: 
\lya\ and \Heiiuv\ emission and the ionising 
properties of high redshift starburst galaxies}

\author{Daniel Schaerer \inst{1}}

\offprints{D. Schaerer, schaerer@ast.obs-mip.fr}

 \institute{Observatoire Midi-Pyr\'en\'ees, Laboratoire d'Astrophysique, UMR
 5572, 14, Av.  E. Belin, F-31400 Toulouse, France }

\date{Received 6 august 2002 / Accepted 18 october 2002}

\titlerunning{\lya\ and \Heiiuv\ emission and ionising properties of starbursts}

\abstract{
%

Using new sets of stellar evolution models at very low metallicities 
($Z = 10^{-7}$, $10^{-5}$) and previously published grids
we examine spectral properties of the ionising continua,
the Lyman-break, and the \lya\ and \Heiiuv\ recombination lines 
in starbursts.
The metallicity dependence of these properties, especially the 
transition from primordial galaxies (Population III) to currently
observed metallicities, is examined for various IMFs and star formation
histories.

For the average properties of starbursts, approximated by a model
with constant star formation, the main findings are:
\begin{itemize}
\item The Lyman continuum flux \Qh\ increases with decreasing metallicity.
For a universal Salpeter IMF from 1--100 \msun\ the enhancement
reaches typically a factor of $\sim$ 3 between solar metallicity and Pop III
objects.
\item While for metallicities $Z \ga 1/50 \zsun$ the amplitude of 
the Lyman-break depends little on $Z$,
a reduction by a factor $\sim$ 2 is found at lower metallicities,
due to the strong increase of the average stellar temperature.
\item Using theoretical models and empirical constraints 
we discuss the expected evolution of the hardness of He$^+$ to H 
ionising photons, \Qrathep, with metallicity and possible uncertainties.
Over the metallicity range $Z=0$ to $\sim 10^{-4}$ the hardness decreases 
from $\log (\Qrathep) \sim -1.4 \ldots -2.3$  
by $\sim$ 1.5--2 or more orders of magnitude, depending strongly on 
the upper mass cut-off of the IMF.
From empirical constraints we derive a hardness $\log (\Qrathep) \sim$
--3.2 to --2.6 for metal-poor starbursts ($1/25 \la Z/\zsun \la 1/4$)
and softer spectra for higher metallicities.
We also provide a simple estimate of the possible impact of hot WR 
like stars on \Qrathep\ at very low metallicities ($Z \la  10^{-4}$).
\item Calibrations for star formation rate determinations from various
recombination lines at all metallicities and for various IMFs are derived.
\end{itemize}

For young bursts the maximum \lya\ equivalent width
is shown to increase strongly with decreasing metallicity from
$W(\lya) \sim$ 250--350 \AA\ at $Z \ga$ 1/50 \zsun\ to 400--850 \AA\
or higher at $Z$ between $10^{-5}$ and 0 (Pop III) for the same
Salpeter IMF.
However, for well known reasons, the \lya\ emission
predicted likely represents an upper limit.

Non-negligible \Heiiuv\ emission due to stellar photoionisation
appears to be limited to very
small metallicities ($\log(Z/\zsun) \la -5.3$) and Population III objects.

The predictions, available on the Web through the CDS and 
at {\tt http://webast.ast.obs-mip.fr/sfr/},
should be useful for a variety of studies regarding
high redshift galaxies, cosmological reionisation, and others.

}
\maketitle
\section{Introduction}
\label{s_intro}

The discovery of numerous high redshift galaxies provides a unique
opportunity to study galaxies in formation in the early Universe.
Most of these galaxies show signs of actively ongoing massive star
formation, as revealed by their overall spectral appearance, by detailed
spectral features, and in many cases by the presence of strong emission
lines.


In fact, since the detection of Lyman-break galaxies at $z \sim$ 2--4 
by colour selection techniques (Steidel \etal\ 1996, review by Stern \& 
Spinrad 1999), \lya\ surveys or other search techniques
have found a large number of objects at higher redshift showing
in most cases intense line emission
(e.g.\  Hu \etal\ 1998, 1999, Kudritzki \etal\ 2000, Rhoads \& Malhotra 2001, 
Malhotra \& Rhoads 2002, Ellis \etal\ 2002, Frye \etal\ 2002, Ajiki \etal\ 2002).
This also includes the most distant galaxy known to date, a lensed
galaxy at $z=6.56$ found through its \lya\ emission (Hu \etal\ 2002).

It is possible that such strong line emitters showing also relatively
little continuum light may represent the earliest stages of galaxy formation,
where small amounts of metals have so far been formed.
Strong ongoing star formation and a small dust content, which can suppress
\lya\ emission, would then explain the high observed \lya\ equivalent 
widths (cf.\ Hu \etal\ 1998).
More striking is the suggestion of Malhotra \& Rhoads (2002)
that the high \lya\ equivalent widths observed in the LALA survey at
$z=4.5$ could, among other explanations, be due to metal-free 
(so-called Population III, hereafter Pop III) objects.
Possibly we are beginning to probe distant chemically little evolved
galaxies, closing the gap between the first (primordial) galaxies
and the high (close to solar) metallicities of massive galaxies
in the local Universe.

To properly study these objects appropriate spectral models are necessary.
They should take into account
all possible metallicities, and also probable systematic changes
of the stellar initial mass function (IMF) at very low metallicity
(Abel \etal\ 1998, Bromm \etal\ 1999, Nakamura \& Umemura 2001,
Hernandez \& Ferrara 2001).
Providing such model calculations is the main aim of the present work.

%
%

Other applications also require an understanding of
how properties like line emission and the ionising fluxes
of starburst behave in the transition between metal-free (Pop III) objects
and metal-poor galaxies with observable counterparts in the local Universe.
This is the case in studies addressing the re-ionisation
history of the Universe (e.g.\ Gnedin 1998, Ciardi \etal\ 2000,
review by Loeb \& Barkana 2000),
especially if account is taken for the simultaneous metal-enrichment
(cf.\ Gnedin \& Ostriker 1997, Ferrara \& Schaerer 2002).
Also, for searches of primordial galaxies, it is of interest
to explore how far extreme properties such as strong \heii\ emission
predicted for Pop III starbursts (Tumlinson \& Shull 2000, Tumlinson
\etal\ 2001, 2002; Oh \etal\ 2001, Bromm \etal\ 2001b, Schaerer 2002) 
are truly limited to zero metallicity.
Our model calculations, including in particular metallicities
$Z=0$, $10^{-7}$, $10^{-5}$,  $4. \, 10^{-4}$ and higher, 
allow, for the first time, such investigations.  

The present paper is structured as follows.
Our models ingredients, including two new sets of stellar evolution tracks
at very low metallicity, are described in Sect.\ \ref{s_models}.
The predicted Lyman continuum fluxes and the properties of the
Lyman-break at all metallicities are presented in Sects.\ \ref{s_qi} 
and \ref{s_lyb} respectively.
In Sect.\ \ref{s_q2} we discuss theoretical predictions and empirical
constraints on the He$^+$ ionising flux and the hardness of the 
ionising spectra of starbursts at various metallicities.
Finally, quantitative predictions for the \lya\ and \Heiiuv\ emission
are given in Sect.\ \ref{s_lya}.
Section \ref{s_conclude} summarises our main conclusions.

\section{Model ingredients}
\label{s_models}

The basic model ingredients are identical to those described in
Schaerer \& Vacca (1998, hereafter SV98) and the Pop III models of
Schaerer (2002a, henceforth S02). A brief summary is provided
subsequently including the new features introduced in the present work.

\subsection{Atmosphere models}
Depending on the metallicity different sets of atmosphere models are used.

For metallicities $Z \le 10^{-5}$ we follow S02 in using the grid of plane 
parallel non-LTE atmospheres computed with the {\em TLUSTY} code of 
Hubeny \& Lanz (1995) for \teff\ $\ge$ 20000 K and 
the plane parallel line blanketed LTE models of Kurucz (1991) with a 
very metal-poor composition ([Fe/H] $= -5.$) otherwise.
Possible uncertainties in the predicted ionising spectra of very metal-poor
stars are discussed in S02.
In comparison with the computations of S02, the use of an extended grid 
of {\em TLUSTY} models in the present paper leads to some small differences 
related to the highest energy range considered here.

For higher metallicities we follow SV98 and adopt for O stars the
{\em CoStar} non-LTE models including stellar winds (Schaerer \& de Koter 1997),
for Wolf-Rayet (WR) stars the pure He spherically expanding 
non-LTE models of Schmutz \etal\ (1992), and Kurucz (1991) models
of appropriate metallicity otherwise.

As discussed in earlier publications (e.g.\ Mihalas 1978, 
Schaerer \& de Koter 1997, Schmutz \etal\ 1992, Tumlinson \& Shull 2000,
Kudritzki 2002)
the inclusion of non-LTE models is the most important ingredient
to obtain accurate predictions for the ionising spectra of 
massive stars.

\subsection{Stellar tracks}
\label{s_tracks}
To explore a wide range of metallicities covering populations from zero
metallicity (Pop III), over low metallicities ($Z \sim 4. \times 10^{-4}$)
such as observed in \hii\ regions in the local Universe, up to solar metallicity
($Z=\zsun=0.02$), we use the following stellar evolution tracks:
{\em 1)} the Pop III tracks covering masses from 1 to 500 \msun\ with no/negligible 
   mass loss compiled in S02 
   (Marigo \etal\ 2001, Feij\'oo 1999, Desjacques 2000),
{\em 2)} new main sequence stellar evolution tracks from 1 to 500 \msun\ computed 
   with the Geneva stellar evolution code for $Z=10^{-7}$,
{\em 3)} non-rotating stellar models from 2 to 60 \msun\ from Meynet \& Maeder (2002) 
   complemented with new calculations for 1 \msun\ and 85--500 \msun\ for $Z=10^{-5}$,
{\em 4)} non-rotating Geneva stellar evolution tracks for masses 0.8--120 \msun\ 
   (up to 150 \msun\ for $Z=0.0004$) from the compilation of Lejeune \& Schaerer (2001)
   for the remaining metallicities $Z=$0.0004, 0.001, 0.004, 0.008, 0.02 ($=$\zsun), 
   and 0.04. For massive stars we have adopted the high mass loss tracks
   in all cases, as these models reproduce best various observational constraints 
   from the local Universe (cf.\ Maeder \& Meynet 1994).   

Our calculations at $Z \le 10^{-5}$ include only the H-burning phase. 
As He-burning is typically less than 10 \% of the main sequence lifetime, and
is generally spent at cooler temperature, neglecting this phase should have little 
or no consequences on our predictions.
A possible exception may be if stars at very low $Z$ are very rapid rotators
(cf.\ Maeder \etal\ 1999) which could suffer from non-negligible rotationally
enhanced mass loss and could therefore become hot WR-like stars
(cf.\ Sect.\ \ref{s_discuss}).

To verify our new computations we have compared several $Z=10^{-5}$ tracks
with the independent calculations of Meynet \& Maeder (2002) done with a strongly
modified version of the Geneva stellar evolution code. 
Good agreement is found regarding the zero age main sequences (ZAMS),
H-burning lifetimes, and the overall appearance of the tracks. 

\begin{figure}
\centerline{\psfig{file=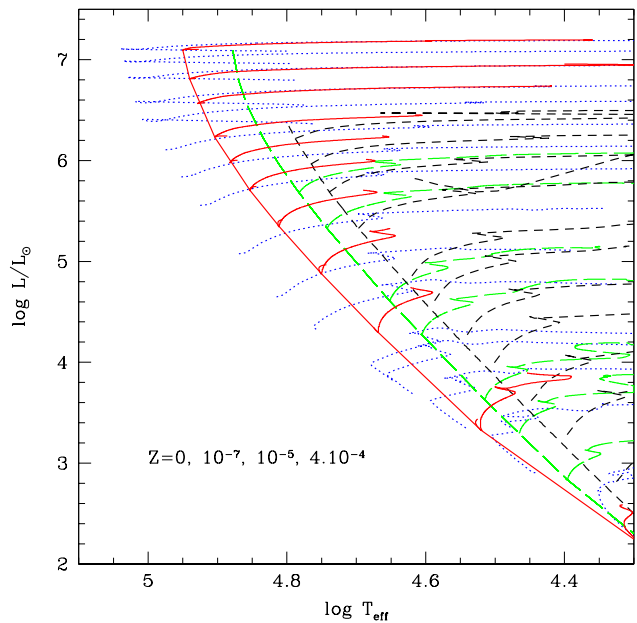,width=8.8cm}}
\caption{HR--diagram showing the main sequence tracks of stars with 
masses $\protect\ga$ 3 to 500 (150 for $Z=0.0004$) \msun\ of metallicities 
$Z=0$ (Pop III, dotted line),
$Z=10^{-7}$ (solid), $Z=10^{-5}$ (long-dashed), and
$Z=4. \, 10^{-4}$ (short-dashed) and their ZAMS.
The source of the tracks is given in Sect.\ \ref{s_models}.
Note the important shift of the ZAMS to high \teff\ from low metallicity
to $Z=0$.}
\label{fig_hrd}
\end{figure}

The HR-diagram showing the main sequence tracks at very low metallicity
is given in Fig.\ \ref{fig_hrd}.
As expected one finds an important shift of the ZAMS and main sequence
toward hotter \teff\ with decreasing $Z$.
For massive stars the trend shown by the low $Z$ tracks is expected 
to continue down to a limiting metallicity $Z_{\rm lim}$ of the order 
of $10^{-9}$, below which the stellar properties essentially
converge to those of metal-free (Pop III) objects.
This is the case, as massive stars ($M \ga$ 50 \msun) with 
$0 \le Z < Z_{\rm lim}$ will rapidly build up a mass fraction $X_{\rm CNO} \sim
5. \, 10^{-10}$ to $2. \, 10^{-9}$ of CNO during pre main sequence
contraction or the early main sequence phase (e.g.\ El Eid \etal\ 1983,
Marigo \etal\ 2001).
A typical value of $Z_{\rm lim} = 10^{-9}$ will be adopted subsequently
for various simplified models fits.

\subsection{Evolutionary synthesis models}
As described in S02 the above stellar atmosphere models and evolutionary 
tracks have been included
in the evolutionary synthesis code of Schaerer \& Vacca (1998).
Using the prescriptions summarised below we compute the predicted
properties of integrated stellar populations at different metallicities
for instantaneous bursts and constant star formation, the two limiting 
cases of star formation histories. 

\begin{table}
\caption{Line emission coefficients $c_l$ in [erg] for Case B, 
$n_e = 100$ cm$^{-3}$ and the different adopted $T_e$.
See S98 and S02 for sources of the atomic data}
\label{tab_lines}
\begin{tabular}[htb]{lllllll}
Line & $c_l$             &  $c_l$             & appropriate $Q_i$  \\
     & $T_e=$30 kK      &  $T_e=$10 kK 
\\ \smallskip
\\ \hline 
\\ \smallskip
\lya\     & $1.04 \times 10^{-11}$ & $1.04 \times 10^{-11}$ & \Qh \\
\Heiiuv\  & $5.67 \times 10^{-12}$ & $6.40 \times 10^{-12}$ & \Qhep \\
\ha\      & $1.21 \times 10^{-12}$ &  $1.37 \times 10^{-12}$ & \Qh \\
\\ \hline 
\end{tabular}
\end{table}

\subsubsection{Nebular emission}
Among other predictions the code in particular computes the
recombination line spectrum including \lya, \Heiiuv, \heii\ $\lambda$3203, 
\hei\ $\lambda$4026, \Hei, \Heiiopt, \hb, \hei\ $\lambda$5016, \hei\ $\lambda$5876, 
and \ha. Case B recombination is assumed for an electron temperature 
of $T_e=30000$ K at $Z \le 10^{-5}$ and $T_e=10000$ K otherwise, 
and a low electron density ($n_e = 100$ cm$^{-3}$).
\lya\ emission is computed assuming a fraction of 0.68 of photons converted
to \lya\ (Spitzer 1978).
The line emission coefficients $c_l$ (defined by Eq.\ \ref{eq_lines})
of interest here are listed in Table \ref{tab_lines}. 

Some differences, e.g.\ due to a more realistic temperature structure
or due to collisional effects on H lines, can be expected between 
the adopted prescriptions and predictions from detailed photoionisation models
(see e.g.\ Stasi\'nska \& Tylenda 1986, Stasi\'nska \& Schaerer 1999).
However, for the scope of the present investigation these effects can
quite safely be neglected.

As shown by S02 the inclusion of nebular continuous emission processes 
is crucial for very metal-poor objects with intense ongoing star formation.
Following S02 we include free-free, free-bound, and H two-photon continuum 
emission assuming $T_e=20000$ K or 10000 K and the above value of $n_e$.

\begin{table}[htb]
\caption{
Summary of IMF model parameters. 
All models assume a Salpeter slope for the IMF}
\label{tab_models}
\begin{tabular}[htb]{llrrr}
Model ID & \mlow\ & \mup\ & $c_M$ 
\\ \smallskip 
\\ \hline
A         &  1   & 100  &  2.55  \\
B         &  1   & 500  &  2.30  \\
C         &  50  & 500  &  14.5  \\
 \hline
\end{tabular}
\end{table}

\begin{figure*}[htb]
\centerline{\psfig{file=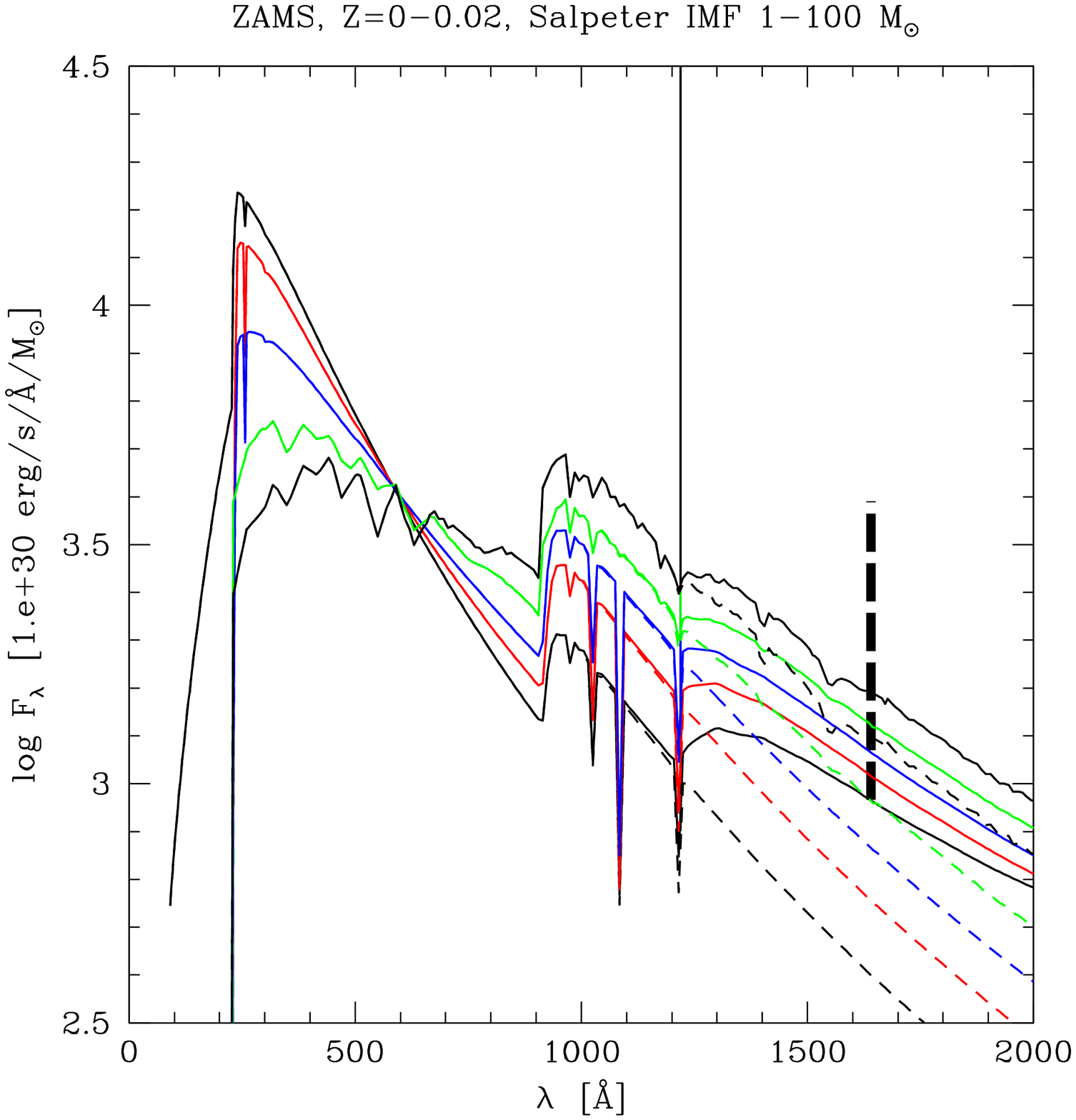,width=8.8cm}
            \psfig{file=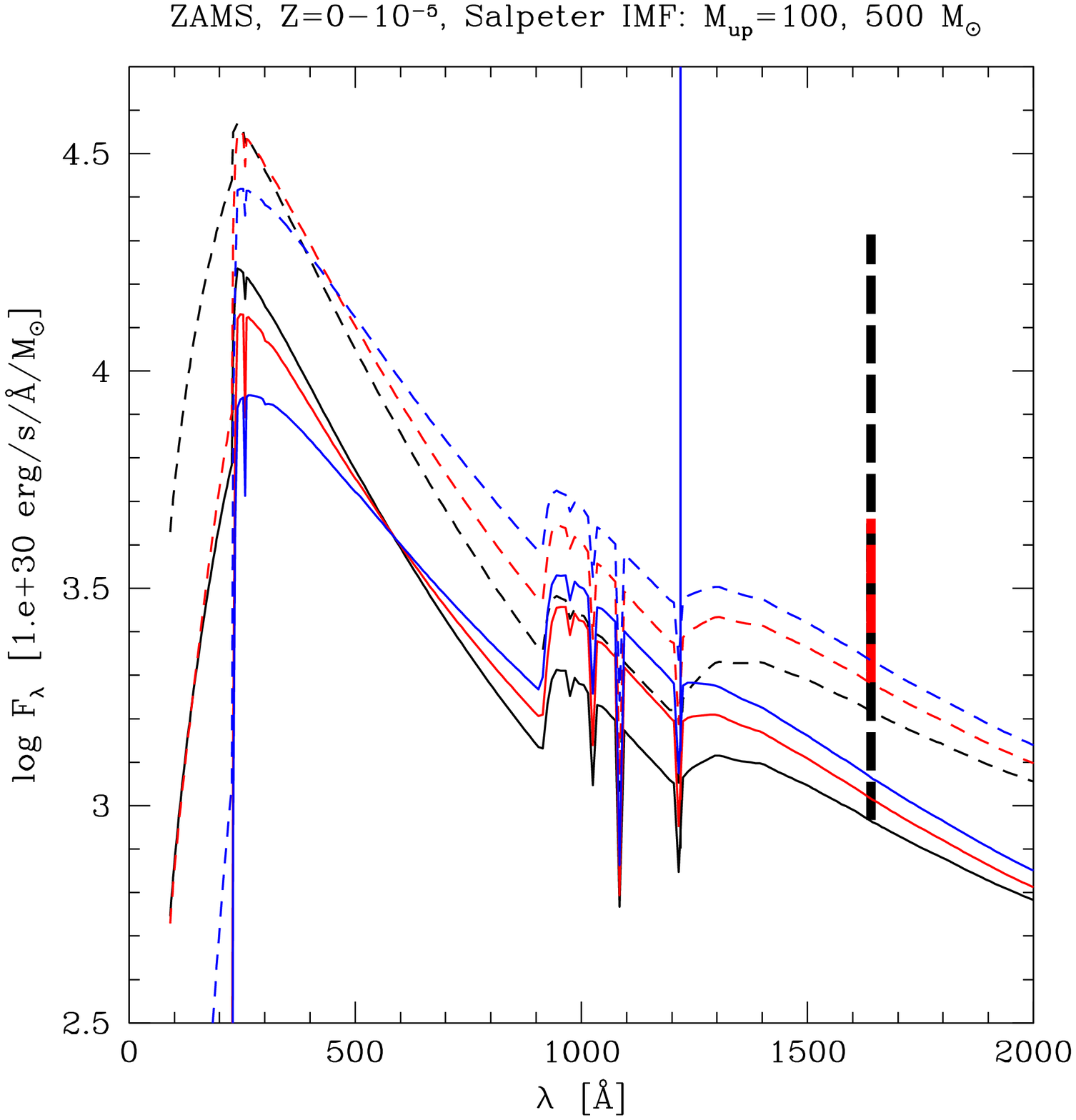,width=8.8cm}}
\caption{Predicted SEDs including \lya\ and \Heiiuv\ emission lines
for zero age main sequence (ZAMS) models at different metallicities.
{\em Left panel:}
Dependence of the SED on metallicity:
The metallicities $Z=$ 0. (Pop III), $10^{-7}$, $10^{-5}$, 0.0004, and 
0.02 (solar) are from top to bottom in the EUV ($\lambda <$ 912 \AA),
and reversed at longer wavelengths. The dashed lines are the pure
stellar emission, the solid lines show the total (stellar $+$ nebular) 
emission. IMF A in all cases.
{\em Right panel:}
Dependence of the SED on \mup\ for metallicities  $Z=$ 0. (Pop III), $10^{-7}$,
and $10^{-5}$. Solid lines show the IMF A, dashed lines
the IMF B. In all cases the total (stellar $+$ nebular) emission is
shown. 
Discussion in text}
\label{fig_seds}
\end{figure*}

\subsubsection{Initial mass function}
In view of our ignorance on the massive star IMF at very low metallicities 
($Z < 4. \times 10^{-4}$) we adopt as in S02 a powerlaw IMF and different
upper and lower mass limits with the aim of 
assessing their impact on the properties of integrated stellar populations.
The main cases modeled here are summarised in Table \ref{tab_models}
\footnote{For comparison with SFR indicators used for ``normal'' galaxies
we also indicate the mass conversion factor $c_M$ expressing
the relative masses between our model IMF and a ``standard'' Salpeter
IMF with \mlow\ $=$ 0.1 and \mup\ $=$ 100 \msun\ used
frequently throughout the literature. $c_M$ is given by
$c_M = \int_{0.1}^{100} M \Phi(M) dm / \int_{\mlow}^{\mup} M \Phi(M) dm$.}.
For all models the IMF slope is taken as the 
Salpeter value ($\alpha =$ 2.35) between the lower and upper 
mass cut-off values \mlow\ and \mup\ respectively.

The model A IMF is a good description of the IMF in observed starbursts
(e.g.\ Leitherer 1998, Schaerer 2002b) down to $\sim$ 1/50 \zsun\ 
($=4. \, 10^{-4}$), the metallicity of I Zw 18 representing the most 
metal-poor galaxy known to date.
It is adopted in all calculations for $Z \ge 4. \, 10^{-4}$.
IMFs B and C, favouring the formation of very massive stars, could be
representative of stellar populations at metallicities 
$Z \la Z_{\rm crit} \approx 10^{-5}$ (e.g.\ Bromm \etal\ 2001a), 
where altered fragmentation properties may form preferentially more 
massive stars (cf.\ Abel \etal\ 1998, Bromm \etal\ 1999, Nakamura \& 
Umemura 2001).
The computations at $Z \le 10^{-5}$ consider all IMF cases (A, B, 
and C). 
Note that our calculations obviously do not apply to cases where
only one or few massive stars form within a pre-galactic halo,
as suggested by the simulations of Abel \etal\ (2002).

\section{Overview of EUV--UV spectra of starbursts at various metallicities}
\label{s_seds}

To illustrate several points discussed in detail below, we plot the 
EUV to UV spectral energy distribution (SED) for selected models at different
metallicities and computed with different IMFs (see Fig.\ \ref{fig_seds}).
Only ZAMS models are shown here for simplicity sake.

The main points apparent from this Figure are:
\begin{itemize}
\item The rapid decrease of the He$^+$ ionising photon flux 
  (i.e.\ flux above 228 \AA) and the associated overall softening of
  the ionising spectrum from Pop III to higher metallicity 
  (cf.\ Sects.\ \ref{s_qi} and \ref{s_q2}).
\item The increase of the Lyman-break amplitude with increasing $Z$
    (Sect.\ \ref{s_lyb}).
\item The rapid decrease of the contribution of nebular continuous emission
    to the UV continuum flux (cf.\ S02).
\item The strong dependence of the ionising flux and in turn also of the
    nebular continuous emission on the upper mass cut-off of the IMF,
    which is considered as a free model parameter for very low
    ($Z \le 10^{-5}$, currently unobserved) metallicities
    (Sects.\ \ref{s_qi} and \ref{s_q2}).
\end{itemize}
The metallicity dependence of the emission lines (strongest in the Pop 
III models) cannot clearly be discerned from this figure.
It is discussed in Sect.\ \ref{s_lya}.

\section{Ionising properties of starbursts at various metallicities}
\label{s_qi}

\begin{figure}
\centerline{\psfig{file=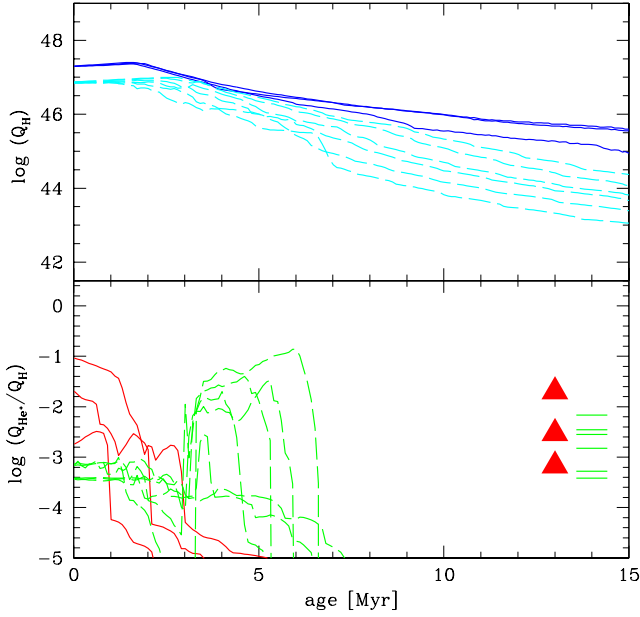,width=8.8cm}}
\caption{Temporal evolution of the H ionising photon flux \Qh\ (upper panel)
and the hardness \Qrathep\ (lower panel) for instantaneous bursts
at all metallicities $Z$ between 0.04 (=2 \zsun) and 0. (Pop III).
The very metal-poor models ($Z= 10^{-5}$, $10^{-7}$ and 0) with the IMF B (1-500 \msun)
are shown as solid lines from top to bottom. The remaining metallicities 
computed for the IMF A (1-100 \msun) are shown with dashed lines. 
Their hardness \Qrathep\ reached for constant star formation at equilibrium
is shown by the solid triangles ($Z \le 10^{-5}$), and the short lines for
larger $Z$.
Predictions for IMF cases not shown here are summarised in
Tables \ref{tab_zams} and  \ref{tab_csf}.
Discussion in text
}
\label{fig_qi}
\end{figure}

\begin{table}[htb]
\caption{ZAMS properties of integrated stellar populations at various metallicities.
All quantities are normalised to a total burst mass of 1 \msun. The logarithm
of the ionising photon fluxes $Q_i$ in units of [photon s$^{-1}$ \msun$^{-1}$] is given.
}
\begin{tabular}{rrrrrrrrrr}
$Z$ & IMF & $\Qh$ &  $\Qhe$ &  $\Qhep$ &  $\Qhtwo$ & \\
    &     &  \multicolumn{4}{c}{[log((photon s$^{-1}$)/(\msun\ yr$^{-1}$))]} \\
\hline \\ 
0.          & A  & 46.98 & 46.75 & 45.54 & 46.22 \\
0.          & B  & 47.29 & 47.10 & 46.26 & 46.40 \\
0.          & C  & 47.98 & 47.80 & 47.05 & 46.96 \\
$10^{-7}$   & A  & 46.94 & 46.65 & 43.45 & 46.36 \\
$10^{-7}$   & B  & 47.30 & 47.06 & 45.61 & 46.55 \\
$10^{-7}$   & C  & 48.01 & 47.78 & 46.39 & 47.14 \\
$10^{-5}$   & A  & 46.90 & 46.55 & 42.39 & 46.44 \\
$10^{-5}$   & B  & 47.30 & 46.99 & 44.56 & 46.64 \\
$10^{-5}$   & C  & 48.02 & 47.73 & 45.35 & 47.24 \\
\\ 
0.0004      &A   & 46.88 & 46.43 & 43.4 & 46.45 \\
0.0004      &A$^\star$& 47.01 & 46.58 & 43.62 & 46.53 \\ 
0.001       & A  & 46.86 & 46.41 & 43.44 & 46.47 \\
0.004       & A  & 46.85 & 46.38 & 43.41 & 46.50 \\
0.008       & A  & 46.84 & 46.35 & 43.67 & 46.52 \\
0.020       & A  & 46.85 & 46.34 & 43.72 & 46.54 \\
0.040       & A  & 46.87 & 46.33 & 43.71 & 46.59 \\
\hline \\
\noalign{\smallskip}
\noalign{
$^\star$: Salpeter IMF with \mup=150 \msun\ and \mlow=1 \msun}
\end{tabular}
\label{tab_zams}
\end{table}

\subsection{Burst models}
\label{s_ib}
The basic quantities describing the ionising spectrum are the
emitted number of H, He, and He$^+$ ionising photons, 
denoted by \Qh, \Qhe\ and \Qhep\ respectively, 
and the hardness \Qrathep\ (\Qrathe) 
tracing the energy range between 54 (24.6) and 13.6 eV.
The predicted temporal evolution of $\Qh$
is shown in Fig.\ \ref{fig_qi} (upper panel)
for all metallicities between $Z=0.$ and $2 \times \zsun$.
For the very low metallicities ($Z \le 10^{-5}$) 
only the models with 
an IMF extending to 500 \msun\ (model B) are shown for clarity sake.
Adopting a larger value of \mup\ affects only the predictions at very
young ages (ages $\la$ 2.5 Myr) due to very short lifetime of these stars.

The predicted $Q_i$ of ZAMS populations (age=0) for all IMF cases
 are listed in Table \ref{tab_zams}.
For completeness with S02 the photon flux in the Lyman-Werner band 
(11.2--13.6 eV) capable to dissociate H$_2$ is also listed. 

The main difference in the Lyman continuum photon output at different
$Z$ is a slower decline of the ionising photon production at low metallicities,
due to the blueward shift of the main sequence.
The temporal evolution of \Qh\ at $Z=10^{-7}$ is essentially undistinguishable
from the Pop III case. 
The larger \Qh\ apparent for $Z \le 10^{-5}$ at ages $\la$ 2.5 Myr
are essentially due to the larger value of \mup\ adopted at very low Z.
The difference at older ages (when stars with masses $>$ 100 \msun\ 
have disappeared) represents the pure metallicity difference.

As can be seen from Table \ref{tab_zams} the \Qh\ production of ZAMS 
populations at different metallicities increases somewhat with decreasing
$Z$; the changes remain fairly small ($\la$ 40 \%) in reasonable
agreement with other estimates (e.g.\ Tumlinson \& Shull 2000).
However, in cases such as constant star formation 
(equivalent to a temporal average) the $Z$-dependence
is more pronounced (cf.\ Sect.\ \ref{s_csf}).

\subsection{Constant star formation}
\label{s_csf}
The main predictions for models with constant star formation at all
metallicities and for all the IMF cases are
listed in Table \ref{tab_csf}. In this case most quantities of interest
here reach rapidly (over timescales $\la$ 6--10 Myr; 
except for the Lyman-break and \Qhtwo\ requiring $\ga$ 200 Myr) an equilibrium
value given in the table, normalised to a star formation rate (SFR) of 
1 \msun\ yr$^{-1}$.
In addition to the ionising photon production $Q_i$ (cols.\ 3-5), and the $H_2$
photodissociating photon flux (\Qhtwo, col.\ 6), we list the average energies
$\overline{E}(\Qh)$ and $\overline{E}(\Qhep)$ of the Lyman continuum photons
and the photons with energies above 54 eV (cols.\ 7 and 8).
These quantities, not further discussed here, are e.g.\ of interest to 
estimate the thermal evolution of the ISM.
Most of the data for $Z=0$, 0.0004, an 0.02 were already given in Table 3 of S02.
Due to the use of a finer grid of atmosphere models at $Z=0$ 
some small changes are found for these models\footnote{$f_{1640}$ and the 
corresponding \Qhep\ are reduced by $\sim$ 9, 4, and 30 \% for IMFs A, B, and C
respectively.}. The values in Table
\ref{tab_csf} supersede those of S02.

As expected from the earlier discussion (see Fig.\ \ref{fig_qi}), 
the Lyman continuum flux \Qh\ shows an increase with decreasing metallicity,
which can be fitted to an accuracy better than 10 \% by
\begin{equation}
  \log(\Qh) = -0.0029 \times (\log(Z) + 9.)^{2.5} + 53.81
\label{eq_qh}
\end{equation}
valid for the Salpeter IMF from 1--100 \msun\ (models A) and 
$Z \in [10^{-9},0.04]$. The limitation to $Z \ge Z_{\rm lim} \approx 10^{-9}$ 
is due to the fact that all metallicities below $Z_{\rm lim}$
are equivalent to the Pop III ($Z=0$) case, as discussed in
Sect.\ \ref{s_tracks}.
Fits for alternate IMFs may be derived from the data in Table \ref{tab_csf}.
 
Overall, while ZAMS Lyman continuum fluxes vary by less than $\la$ 40 \%
over the entire metallicity range for IMF A (Table \ref{tab_zams}), 
the ionising output at SFR$=const$ shows an increase of an factor 
$\sim$ 1.9 (2.8) between solar and 1/50 \zsun\ (zero metallicity).
Even larger increases are of course predicted
in the case of IMFs extending to higher masses (models B and C).

\begin{figure}[htb]
\centerline{\psfig{file=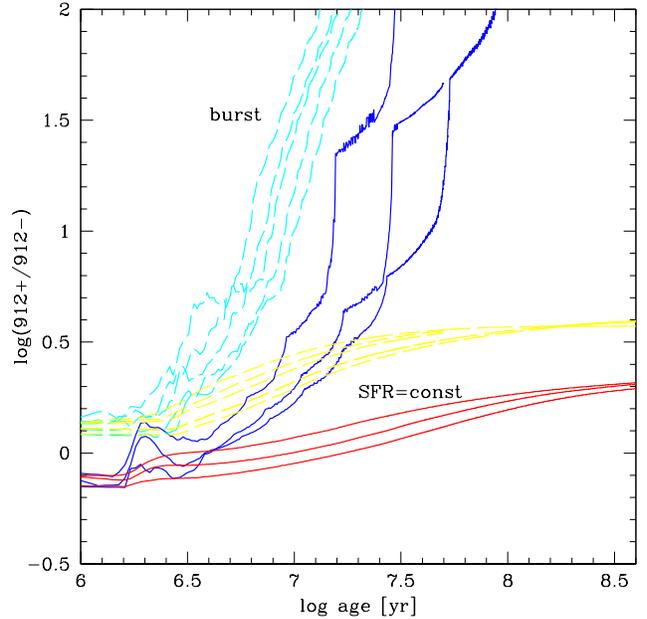,width=8.8cm}}
\caption{Temporal evolution of the Lyman-break for instantaneous bursts
and constant star formation at all metallicities $0 \le Z \le \zsun$.
Dashed lines show the predictions for $Z \ge 4. \,10^{-4}$,
solid lines are for $Z=0$ (Pop III), $Z=10^{-7}$, and $10^{-5}$
(from bottom to top).
The behaviour of the very low $Z$ burst models at ages $\log t \sim$ 
7.2, 7.5, 7.7 are artifacts due to the limited numerical resolution in 
tracks.
Note the small sensitivity of the Lyman-break for SFR$=const$ at
metallicities $Z \ge 4. \, 10^{-4}$ and the important decrease
at lower metallicities.
}
\label{fig_lyb}
\end{figure}

\begin{table*}[htb]
\caption{Predictions for the case of constant star formation models
(equilibrium values) with different IMFs (col.\ 2) at all
metallicities (col.\ 1).
Given are the ionising photon fluxes $Q_i$ (cols.\ 3--6), their average energies
$\overline{E}$ (cols.\ 7, 8), and the proportionality constants $f_l$ between
recombination line luminosities and the SFR (cf.\ Eq.\ \ref{eq_lines}, 
cols.\ 9--11).
The logarithm of the ionising photon fluxes
$Q_i$ is given in units of [photon s$^{-1}$ \msun$^{-1}$].
}
\begin{tabular}{rrrrrrrrrrrrrrrrrr}
$Z$ & IMF & $\Qh$ &  $\Qhe$ &  $\Qhep$ &  $\Qhtwo$ &   
    $\overline{E}(\Qh)$ & $\overline{E}(\Qhep)$ & $f_{\rm Ly-\alpha}$ & $f_{\ha}$ & $f_{1640}$\\
    &     &  \multicolumn{4}{c}{[log((photon s$^{-1}$)/(\msun\ yr$^{-1}$))]}            
          &  \multicolumn{2}{c}{[eV]} 
          & \multicolumn{3}{c}{[erg s$^{-1}$]}  \\
\smallskip 
\\ \hline  
0.          & A  & 53.81 & 53.50 & 51.49 & 53.57 & 26.61 & 66.08 & 6.80e+42 & 7.91e+41 & 1.74e+40 \\
0.          & B  & 53.93 & 53.64 & 52.23 & 53.57 & 27.78 & 68.15 & 8.86e+42 & 1.03e+42 & 9.66e+40 \\
0.          & C  & 54.44 & 54.19 & 53.03 & 53.65 & 29.60 & 68.22 & 2.85e+43 & 3.32e+42 & 6.01e+41 \\
$10^{-7}$   & A  & 53.80 & 53.43 & 50.39 & 53.67 & 25.18 & 61.68 & 6.59e+42 & 7.67e+41 & 1.40e+39 \\
$10^{-7}$   & B  & 53.95 & 53.61 & 51.42 & 53.69 & 25.99 & 64.86 & 9.26e+42 & 1.08e+42 & 1.49e+40 \\
$10^{-7}$   & C  & 54.51 & 54.21 & 52.21 & 53.84 & 27.16 & 64.80 & 3.34e+43 & 3.89e+42 & 9.29e+40 \\
$10^{-5}$   & A  & 53.70 & 53.26 & 48.71 & 53.65 & 23.95 & 59.62 & 5.15e+42 & 5.99e+41 & 2.91e+37 \\
$10^{-5}$   & B  & 53.88 & 53.48 & 50.71 & 53.69 & 24.74 & 61.77 & 7.82e+42 & 9.10e+41 & 2.88e+39 \\
$10^{-5}$   & C  & 54.49 & 54.13 & 51.50 & 53.97 & 25.61 & 61.77 & 3.20e+43 & 3.73e+42 & 1.82e+40 \\
\\ 
0.0004      &A   & 53.63 & 53.10 & {\em 50.08$^\dagger$} & 53.74 & 21.62 & {\em 52.60$^\dagger$} & 4.38e+42 & 5.10e+41 & {\em 6.79e+38$^\dagger$} \\
0.0004 &A$^\star$& 53.70 & 53.20 & {\em 50.42$^\dagger$} & 53.75 & 21.96 & {\em 61.54$^\dagger$} & 5.22e+42 & 6.08e+41 & {\em 1.50e+39$^\dagger$} \\ 
0.001       & A  & 53.59 & 53.04 & {\em 50.17$^\dagger$} & 53.72 & 21.47 & {\em 60.38$^\dagger$} & 4.01e+42 & 4.67e+41 & {\em 8.39e+38$^\dagger$} \\
0.004       & A  & 53.50 & 52.93 &                 $^\P$ & 53.67 & 21.27 &                 $^\P$ & 3.37e+42 & 4.36e+41 & $^\P$ \\
0.008       & A  & 53.44 & 52.83 &                 $^\P$ & 53.63 & 20.90 &                 $^\P$ & 2.89e+42 & 3.73e+41 & $^\P$ \\
0.020       & A  & 53.36 & 52.75 &                 $^\P$ & 53.56 & 20.84 &                 $^\P$ & 2.44e+42 & 3.16e+41 & $^\P$ \\
0.040       & A  & 53.28 & 52.65 &                 $^\P$ & 53.49 & 20.88 &                 $^\P$ & 2.00e+42 & 2.59e+41 & $^\P$ \\
\hline \\                                                                                            
\noalign{\smallskip}
\noalign{
$^\star$: Salpeter IMF with \mup=150 \msun\ and \mlow=1 \msun}
\noalign{$^\dagger$: uncertain predictions for reasons discussed in Sect.\ \ref{s_q2} }
\noalign{$^\P$: no data provided, as predictions uncertain and strongly overestimated (cf.\ Sect.\ \ref{s_q2}) }
\end{tabular}                                                                                        
\label{tab_csf}
\end{table*}

\section{Lyman-break predictions at various metallicities}
\label{s_lyb}
Quantitative predictions of the amplitude of the Lyman-break, i.e.\ the relative
flux above and below the Lyman edge are of importance for example
to estimate the escape fraction $f_{\rm esc}$ of ionising photons out of their
hosts. Such measurements are possible in relatively nearby
($z \ga$ 0.015) and high redshift galaxies (e.g.\ Leitherer \etal\ 1995,
Deharveng \etal\ 1997, 2001, Steidel \etal\ 2001).

Different working definitions of the Lyman-break exist in the literature.
For simplicity we chose the definition adopted in {\em Starburst99}
(Leitherer \etal\ 1999), quantifying the break amplitude by 912$^+$/912$^-$, where
912$^+$ is the average flux in $F_\lambda$ units over the interval
1080--1120 \AA, and 912$^-$ the average over 870--900 \AA.
A word of caution is appropriate if wavelengths $\lambda \ga \lambda(\lya)$ 
are used as reference for the red side of the Lyman-break.
In this case the continuum may, for very metal-poor objects, include a 
non negligible contribution from nebular continuous emission (see S02) 
which itself also depends on the escape fraction of Lyman continuum photons.
A detailed SED fit should then be undertaken to quantify $f_{\rm esc}$.

The temporal evolution of the Lyman-break predicted by the models
at metallicities $0 \le Z \le \zsun$ is shown in Fig.\ \ref{fig_lyb}
for instantaneous bursts and constant star formation.
The main finding apparent here is the overall reduction of the
Lyman-break at metallicities $Z \la 10^{-4 \ldots -5}$, e.g.\
by 0.2--0.3 dex at very young ages,
for identical IMFs or larger values of \mup.
\footnote{IMFs A and B yield Lyman-break amplitudes differing by less than $\sim$ 0.1 dex.}
This is due to the decrease of the break in individual stars with
increasing \teff\ and to the shift of the main sequence towards hotter temperatures.
%
At higher metallicity ($Z \ga 4. \, 10^{-4}$) this temperature/metallicity
dependence is not important, and e.g.\ the predicted values
at SFR$=const$ (near equilibrium at $t \ga 10^8$ yr) vary by less 
than 10 \% around  $\log(912^+/912^-) \sim$ 0.58. 
In contrast for very low metallicities one has for the same case 
$\log(912^+/912^-) \sim$ 0.3--0.4 for the IMFs A and B.

Note that, as expected, at $Z \ge 10^{-3}$ our predictions at young ages 
are somewhat ($\sim$ 0.1 dex) smaller than the {\em Starburst99} models 
for identical $Z$ and IMF, due to the inclusion of non-LTE O star model 
atmospheres in the present computations.
For constant star formation this difference becomes, however, smaller.
We also compared our calculations to the models of Smith \etal\ (2002;
Norris, private communication) including more sophisticated non-LTE 
stellar atmospheres. For metallicities $0.001 \le Z \le \zsun$ and SFR$=const$
their Lyman-break predictions show an additional but small ($\sim$ 10-20 \%) 
reduction.

\section{The \Qrathep\ hardness of starbursts at various metallicities}
\label{s_q2}

\subsection{Predictions at metallicities $Z=$ 0 to \zsun}
\subsubsection{Burst models}
The time evolution of the predicted hardness \Qrathep\ 
is shown in the lower panel of Fig.\ \ref{fig_qi}.
The corresponding hardness reached at equilibrium in the case of constant
star formation (cf.\ below), plotted on the right, illustrates
the non negligible difference with the hardness derived from a simple ZAMS 
population neglecting stellar evolution effects (cf.\ S02).

Note that predicted quantities such as \Qrathep, rely obviously strongly on the 
adopted value of \mup. For the very low $Z$ this quantitative 
dependence can be estimated from the tabulated ZAMS properties (Table
\ref{tab_zams}).

As expected from the strong decrease of the stellar temperatures
with increasing metallicity (cf.\ Fig.\ \ref{fig_hrd})
both the maximum hardness (at age=0) and the overall \Qrathep\
decreases for $Z$ between 0 and $10^{-5}$.
The typical timescale for a decrease of \Qrathep\ by $\sim$ 2 dex in a burst
is driven by the redward stellar evolution, and is short 
($\sim$ 2--3 Myr), with obvious potential implications
for the detection of sources with very hard spectra (cf.\ Sect.\ 
\ref{s_lya}).
Possible additional sources of He$^+$ ionising photons not included here
are discussed in Sect.\ \ref{s_discuss}.


At higher metallicities ($Z \ga\ 4. \times 10^{-4}$) the 
present models predict a re-increase of \Qrathep\ after $\ga$ 3--4 Myr,
due to presence of WR stars, among which a fraction is found
at high temperatures (cf.\ Schmutz \etal\ 1992, SV98).
Albeit with minor quantitative differences, a qualitatively similar
behaviour is predicted by the {\em Starburst99} models based
on very similar input physics (Leitherer \etal\ 1999).
However, these predictions depend especially on the procedure
adopted to link stellar tracks with atmospheres in WR phases
with strong winds, and on the neglect of line blanketing
in the adopted WR model atmospheres.
The reality and the extent of such a trend remains therefore
questionable, especially at the largest metallicities
(cf.\ review of Schaerer 2000).
Indeed using recent line blanketed O and WR atmospheres and 
different prescriptions to connect the interior and atmosphere
models Smith \etal\ (2002) find a considerably softer
spectrum -- i.e.\ reduced  \Qrathep\ -- before and during the WR
phase.
To circumvent this theoretical uncertainty we will subsequently
derive an empirical estimate of the hardness
at metallicities $Z \ga 4. \times 10^{-4}$ (Sect.\ \ref{s_emp}).

\begin{figure}[tb]
\centerline{\psfig{file=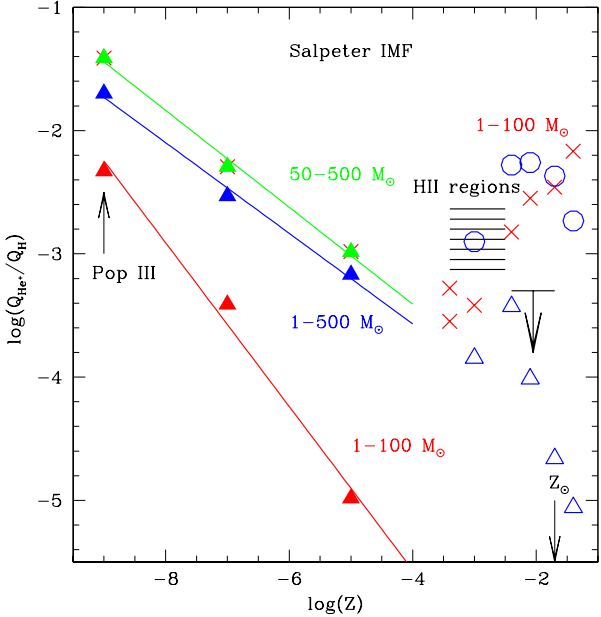,width=8.8cm}}
\caption{Hardness \Qrathep\ of the He$^+$ ionising flux for constant star 
formation as a function of metallicity (in mass fraction) 
for all models given in Table \ref{tab_csf}.
At metallicities above $Z \ge 4. \, 10^{-4}$ the predictions from
our models (crosses), as well as those of Leitherer \etal\ (1999, 
open circles), and Smith \etal\ (2002, open triangles) are plotted.
The shaded area and the upper limit (at higher $Z$) indicates 
the range of the empirical hardness estimated from 
\hii\ region observations (see Sect.\ \ref{s_emp}).
At very low $Z$ solid lines show the fits to the data given 
by Eq.\ \ref{eq_q2}.
Discussion in text}
\label{fig_q2rat}
\end{figure}

\begin{figure}[tb]
\centerline{\psfig{file=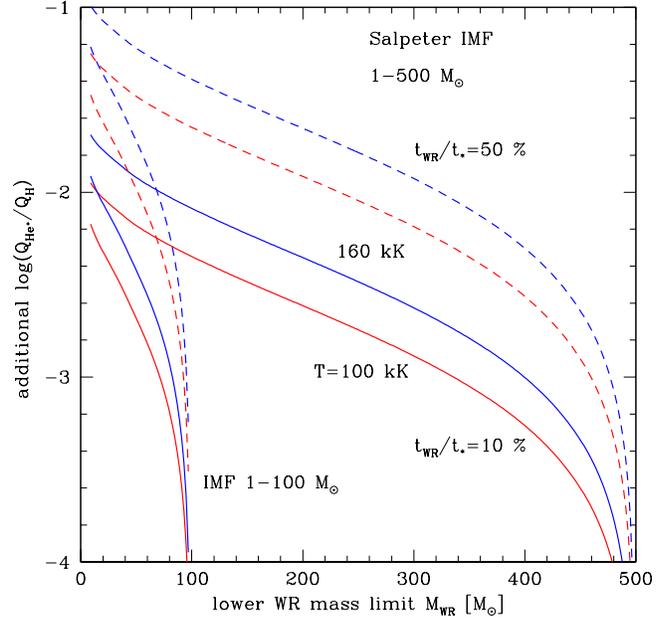,width=8.8cm}}
\caption{Additional contribution $\Qhep^\star/\Qh$ of putative
hot WR-like stars to the total He$^+$ hardness of the ionising flux
$\Qhep/\Qh$ (cf.\ Eq.\ \ref{eq_q2est}) for SFR$=const$ as a function
of the minimum mass $M_{\rm WR}$.
Shown are the cases for $t_{\rm WR}/t_\star=$ 10 and 50 \% (solid and
dashed curves), temperatures $T=$ 100 and 160 kK (lower and upper curve
respectively), and \mup=500 (100) \msun\ extending over the full plot
(lower left only).
}
\label{fig_est}
\end{figure}

\subsubsection{Constant star formation}
For the case of constant star formation at equilibrium
the metallicity dependence of the hardness \Qrathep\ of the ionising flux
is shown in Fig.\ \ref{fig_q2rat} for all the IMFs considered.
As apparent, for the very metal-poor cases ($Z \la 10^{-4}$) \Qrathep\
can be well fitted by 
\begin{equation}
\log(\Qrathep) = a \times \log(Z) + b 
\label{eq_q2}
\end{equation}
with the numerical coefficients listed in Table \ref{tab_fit}, again
taking into account that all metallicities 
$Z \le Z_{\rm lim} \approx 10^{-9}$ are equivalent.

At higher metallicities --- in the $Z$ range of known objects ---
the theoretical predictions for \Qrathep\ are probably less clear, 
due to possible presence of hot WR stars, difficulties in their modeling 
(cf.\ above), and the neglect of non-stellar emission
processes which could contribute especially to \Qhep.
Indeed as shown in Fig.\ \ref{fig_q2rat} rather important differences are
obtained between various evolutionary synthesis models (SV98, {\em Starburst99}
of Leitherer \etal\ 1999, and the latest computations of Smith \etal\ 2002).
In the metallicity range $Z \sim 8. \, 10^{-4}$ to $5. \, 10^{-3}$ our empirical
estimate of \Qrathep\ (Sect.\ \ref{s_emp}), also shown on this Fig., 
is likely more reliable than the models. 
At still larger $Z$ the average ionising spectrum of starbursts
should be softer, as indicated by the tentative empirical upper limit
and predicted by the Smith \etal\ (2002) models.

\subsection{Uncertainties on the He$^+$ ionising flux 
  at very low metallicities ($Z \protect\la 10^{-4}$)}
\label{s_discuss}
In contrast to the Lyman continuum flux \Qh, the He$^+$ ionising flux 
(and in general spectral features at high energy) show 
a very strong dependence on the stellar 
temperature in the \teff\ range $\sim$ 50--100 kK typical of very low 
metallicity stars (Fig.\ 2 in S02).
Therefore their prediction is naturally 
more sensitive to even small modifications
of the exact stellar \teff\ or evolutionary scenario.


For example, one may wonder how reliable the above predictions
of the metallicity dependence of \Qrathep\ 
(Fig.\ \ref{fig_q2rat}) are at $Z \la 10^{-4}$,
where presently no observational constraints are available.
In fact, studies of massive stars in the Local Group suggest
that their average rotation rates increase towards low $Z$
(Maeder \etal\ 1999), which -- when combined with their 
increased compactness -- can lead to non-negligible mass loss
despite the low metallicity (Maeder \& Meynet 2000, Meynet \& Maeder 
2002).
If this effect is large enough, one could imagine that fast rotators 
could loose sufficient mass to follow a WR star like evolution
leading possibly to a He/C/O core at temperatures $T \ga$ 100 kK,
a scenario known for metal-rich massive WR (e.g.\ Maeder \& Meynet 1988).
Despite high rotational velocities, the detailed calculations of 
Meynet \& Maeder (2002) for $Z=10^{-5}$ do not show important alterations
of the evolution for stars with $M \le$ 60 \msun.
Exploratory calculations of Marigo \etal\ (2002) for Pop III stars
treating in an simplified manner the effects of rotation on stellar
mass loss find such a scenario for stars with initial mass $\ga$ 750 \msun.

Quantitatively the effect of such a putative hot ``WR-like'' population on the hardness
of the ionising flux can be estimated {\em for the case of constant star formation
only} in the following way.
Suppose that stars of given initial mass $M \ge M_{\rm WR}$ spend this phase at
constant luminosity $L$ and (hot) temperature $T$
during a constant fraction  $f_{\rm WR}=(t_{\rm WR}/t_\star)$ 
of their lifetime $t_\star$.
Assuming their winds are optically thin at $\ge$ 54 eV,
the He$^+$ flux in this phase is then
\begin{equation}
\Qhep^\star(M) = \frac{L(M)}{\sigma T^4} \qhep^\star(T),
\label{eq_q2s}
\end{equation}
where $\qhep^\star$ is the photon flux per unit surface area (cf.\ Fig.\ 2 in S02).
The hardness $(\Qhep/\Qh)^\prime$ of a population including both the
``normal'' (main sequence) population and the additional hot population
can be approximated by (for SFR$=const$)
\begin{equation}
\left(\frac{\Qhep}{\Qh}\right)^\prime \approx
        \frac{\Qhep}{\Qh} \left(1-f_{\rm WR}\right) + \frac{\Qhep^\star}{\Qh} f_{\rm WR},
\label{eq_q2q0}
\end{equation}
assuming small changes in the total Lyman continuum photon output \Qh.
Here the first term stands for the unperturbed ``normal'' hardness
shown in Fig.\ \ref{fig_q2rat}. The second term describes the additional
contribution due to hot WR-like objects, which can be expressed as 
\begin{equation}
        \frac{\Qhep^\star}{\Qh} = 
        \frac{\int_{M_{\rm WR}}^{\mup} t_\star(M)
                \Qhep^\star(M) \Phi(M) dM}
             {\int_{\mlow}^{\mup} t_\star(M) \Qhave(M) \Phi(M) dM},
\label{eq_q2star}
\end{equation}
where 
$\Qhave(M)$ is the lifetime average of the Lyman continuum production,
and $ \Phi(M)$ is the IMF. 
The quantities $t_\star(M)$ and $\Qhave(M)$ are taken from S02, and
we assume as first approximation a luminosity $L(M)$ corresponding to 
the ZAMS\footnote{The results discussed below are fairly insensitive
to the exact (low) metallicity adopted for these quantities.}. 
The hardness contribution $\Qhep^\star/\Qh$ is then computed
assuming temperatures $T=$ 100--160 kK as expected from stellar evolution
models, 
and durations $f_{\rm WR}=(t_{\rm WR}/t_\star)$ of 0.1 or 0.5.
The result is plotted for the IMFs A and B (\mup=100 and 500 \msun) 
in Fig.\ \ref{fig_est} as a 
function of the minimum mass $M_{\rm WR}$, above which all stars 
are assumed to reach this hot ``WR'' phase.

The case of $(t_{\rm WR}/t_\star)=$ 50 \%, which would require
very strong mass loss already during the main sequence or a nearly homogeneous
evolution leading early to a blueward evolution, appears extremely unlikely
and is shown here to mimic the ``strong mass loss'' models adopted 
in the Pop III models of S02.
For a hot phase of a duration typical of the post main sequence evolution
($\sim$ 10 \% of total lifetime) Fig.\ \ref{fig_est} shows that such
putative ``hot WR'' could in the ``best'' case contribute an additional hardness 
$\Qhep^\star/\Qh$ of the order of $10^{-2 \ldots -3}$, 
comparable to the hardness of normal stellar populations with 
metallicities $Z \la 10^{-4}$ (cf.\ Fig.\ \ref{fig_q2rat}).
To examine how realistic such cases may be, will require a detailed
understanding of the coupled processes of stellar mass loss, rotation,
and internal mixing.
At present the available limits are $M_{\rm WR} >$ 60 \msun\ at 
$Z= 10^{-5}$  and $(t_{\rm WR}/t_\star) \la$ 10 \% 
from the rotating stellar models of Meynet \& Maeder (2002),
and $M_{\rm WR} \ga$ 750 \msun\ at $Z=0$ from the simplified models
of Marigo \etal\ (2002).

Although the above exercise shows that at very low metallicity
($Z \la 10^{-5}$) the hardness \Qrathep\ due to stellar sources could 
be higher than shown in Fig.\ \ref{fig_q2rat}, it seems that such scenarios
are quite unlikely.
%
%
If star formation takes place on much longer time scales, and
massive stars would not form (or in much smaller quantities),
hot planetary nebulae could also be a source of hard ionising photons,
as illustrated by the scenario of Shioya \etal\ (2002).
In any case, a major uncertainty stems from our limited knowledge
of the IMF at very low metallicities.

\begin{table}
\caption{Fit coefficients for Eq.\ \protect\ref{eq_q2}}
\begin{tabular}[htb]{lllrr}
IMF  & a                 & b 
\\ \smallskip 
\\ \hline
A    & -0.66 $\pm$ 0.071 & -8.22 $\pm$ 0.51 \\
B    & -0.37 $\pm$ 0.028 & -5.04 $\pm$ 0.20 \\
C    & -0.39 $\pm$ 0.028 & -4.98 $\pm$ 0.20 \\
\hline
\end{tabular}
\label{tab_fit}
\end{table}

\subsection{Empirical constraints on the He$^+$ ionising flux of
starbursts at $Z \protect\ga 4. \times 10^{-4}$}
\label{s_emp}

Spectroscopic observations of extra-galactic giant \hii\ 
regions probing \heii\ recombinations lines can yield empirical 
information on the ``average'' hardness \Qrathep\ of starbursts.
Indeed it is well known that a fairly large fraction of metal-poor
\hii\ regions show the presence of nebular \Heiiopt\ emission
indicative of a hard ionising spectrum (see e.g.\ Guseva \etal\ 2000,
compilation of Schaerer \etal\ 1999).
A complete explanation of the origin of the required high energy 
photons (shocks, X-rays, WR stars) remains to be found
(e.g.\ Garnett \etal\ 1991; Schaerer 1996, 1998;
Guseva \etal\ 2000, Izotov \etal, 2001, Stasi\'nska \& Izotov 2002).

The largest sample of high quality data is that of Izotov and 
collaborators (cf. Guseva \etal\ 2000 and references therein), which
shows \Heiiopt\ detections with typical relative intensities of 
$I(4686)/I(\hb) \sim$ 1--2 \%.

From such a sample we may estimate an average hardness from
\begin{equation}
\frac{\Qhep}{\Qh} = \frac{I(4686)}{I(\hb)} \times a \times f_{\rm detect} 
                  \times \frac{t_{\rm detect}}{t_{\rm total}},
\label{eq_q2est}
\end{equation}
where $a$ ($=$0.47 (0.63) for $T_e=$ 10000 (30000) K) translates the relative 
line emissivities,
$f_{\rm detect}$ is the fraction of objects showing nebular \heii\ 
among the spectra of sufficient quality to allow detections down to
this intensity level,
$t_{\rm detect}$ is the age range covered by the observed \hii\ regions, and
$t_{\rm total}$ is the total lifetime of \hii\ regions.
Inspection of the data of Izotov yields $f_{\rm detect} \sim$ 1/3--1/2
for metallicities $Z \in  [8. \, 10^{-4}, 4.8 \, 10^{-3}]$ and possibly
even a larger fraction at the lowest observed metallicities.
(Stasi\'nska \& Izotov 2002).
One has $t_{\rm total} \sim$ 10 Myr, 
the typical \hii\ region lifetime or
the time over which bursts show detectable line emission.
However, it is now well established that \hii\ region samples suffer from 
selection biases leading to an absence of advanced bursts, or 
in other words a preferential selection of
bursts with ages younger than $\la$ 4--5 Myr
(e.g.\ Bresolin \etal\ 1999, Raimann \etal\ 2000, Moy \etal\ 2001,
Stasi\'nska \etal\ 2001).
{\em For this range of values and $a=0.47$ one obtains an estimate of 
$\log(\Qrathep) \sim$ --3.2 to --2.6.} 
Postulating the extreme case that additionally the first 2 Myr of the 
\hii\ phase are also not detected in the optical\footnote{The 
well known absence of
\hii\ regions with very large \hb\ equivalent widths could be taken as
an indications for such a scenario.} one has a minimum value of 
$t_{\rm detect} \sim$ 1 Myr, yielding a lower limit of 
$\log(\Qrathep) \ga$ --3.5.
In principle this average value could be even larger 
as the (conservative) assumption of $t_{\rm detect}/t_{\rm total} < 1$
implies that older (non-detected) \hii\ regions have no He$^+$ ionising flux.

This estimate is obviously independent of the nature of the hard (He$^+$) 
ionising radiation.
By construction Eq.\ \ref{eq_q2est} provides an estimate of the average
\Qrathep\ expected in objects with constant ongoing star formation for
metallicities $Z \sim 8. \, 10^{-4}$ to $5. \, 10^{-3}$.
The near absence of nebular \heii\ detections in \hii\ regions
at higher metallicity (cf.\ Schaerer 1998, Guseva \etal\ 2000) indicates
softer spectra. However, it is difficult to establish a firm upper limit 
on \Qrathep\ for $Z \ga 5. \, 10^{-3}$. We here retain $\log(\Qrathep)
\ll -3.3$ as a tentative limit.


\subsection{Overall behaviour of \Qrathep\ with metallicity}
\label{s_q2short}

In short, from the considerations above, we find the following two cases 
for the most plausible metallicity dependence of 
the average \Qrathep\ hardness ratio of starbursts with metallicity
(see Fig.\ \ref{fig_q2rat}).

1) If a universal Salpeter like IMF with a ``normal'' upper mass limit 
of $\sim$ 100 \msun\ prevails for all metallicities
the hardness decreases by more than 2 orders of magnitude from 
$Z=0$ to $\sim 10^{-4}$, re-increases thereafter (up to $Z \la 5.\ 10^{-3}$,
in metal-poor starbursts) 
to a level $\sim$ 2 to 10 times smaller than that of Pop III objects,
and decreases again to low levels for higher metallicities.

2) If very massive stars are favoured at metallicities $Z \la  10^{-4}$,
the hardness \Qrathep\ of Pop III objects is considerably enhanced (corresponding
to a powerlaw spectrum with spectral index $\alpha \sim$ 2.3--2.8 in $F_\nu$), 
then decreases down to levels somewhat smaller or comparable to
that of metal-poor starbursts, before decreasing further to levels at least
two orders of magnitude softer than at zero metallicity.

\begin{figure*}[tb]
\centerline{\psfig{file=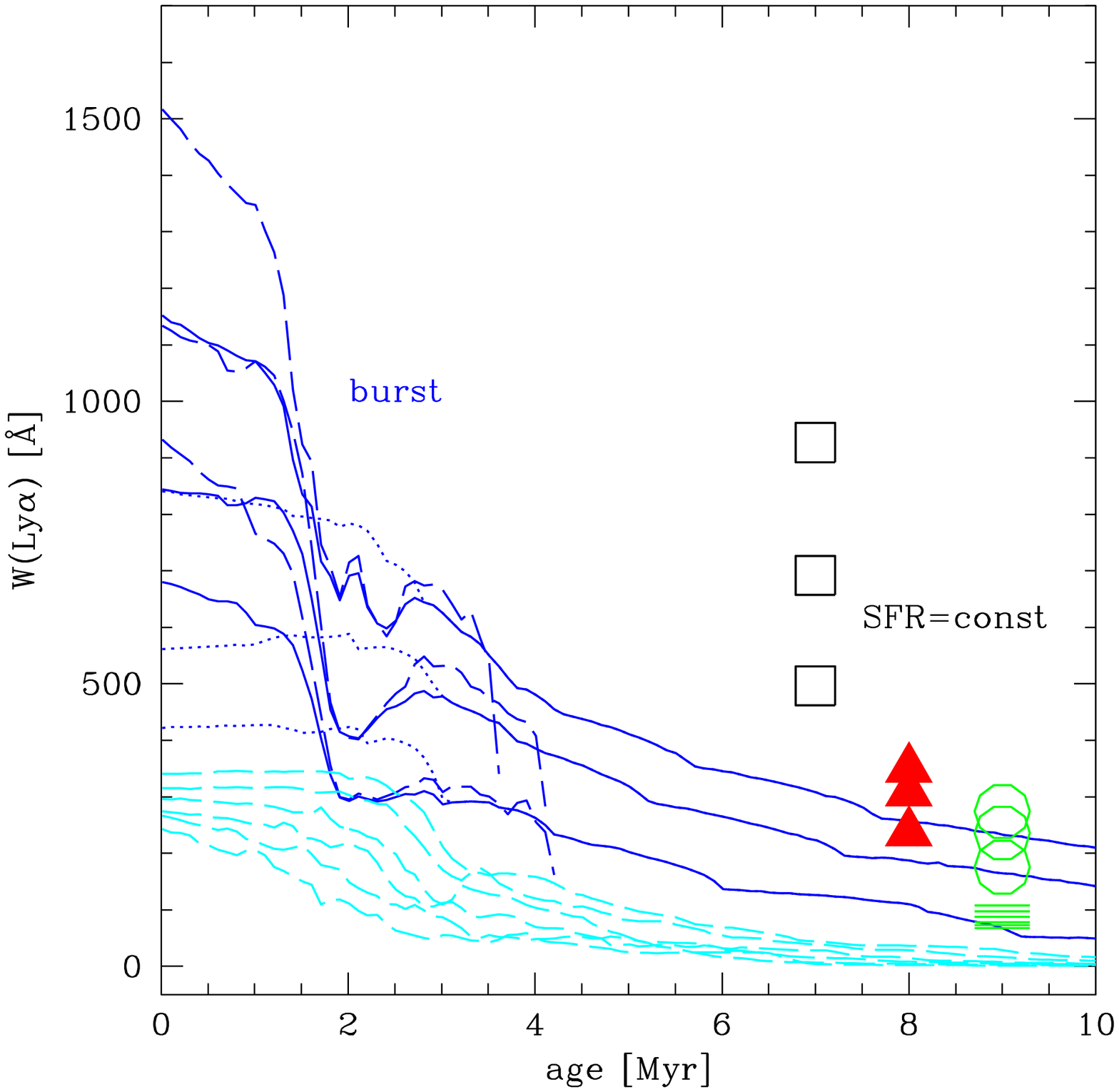,width=8.8cm}
            \psfig{file=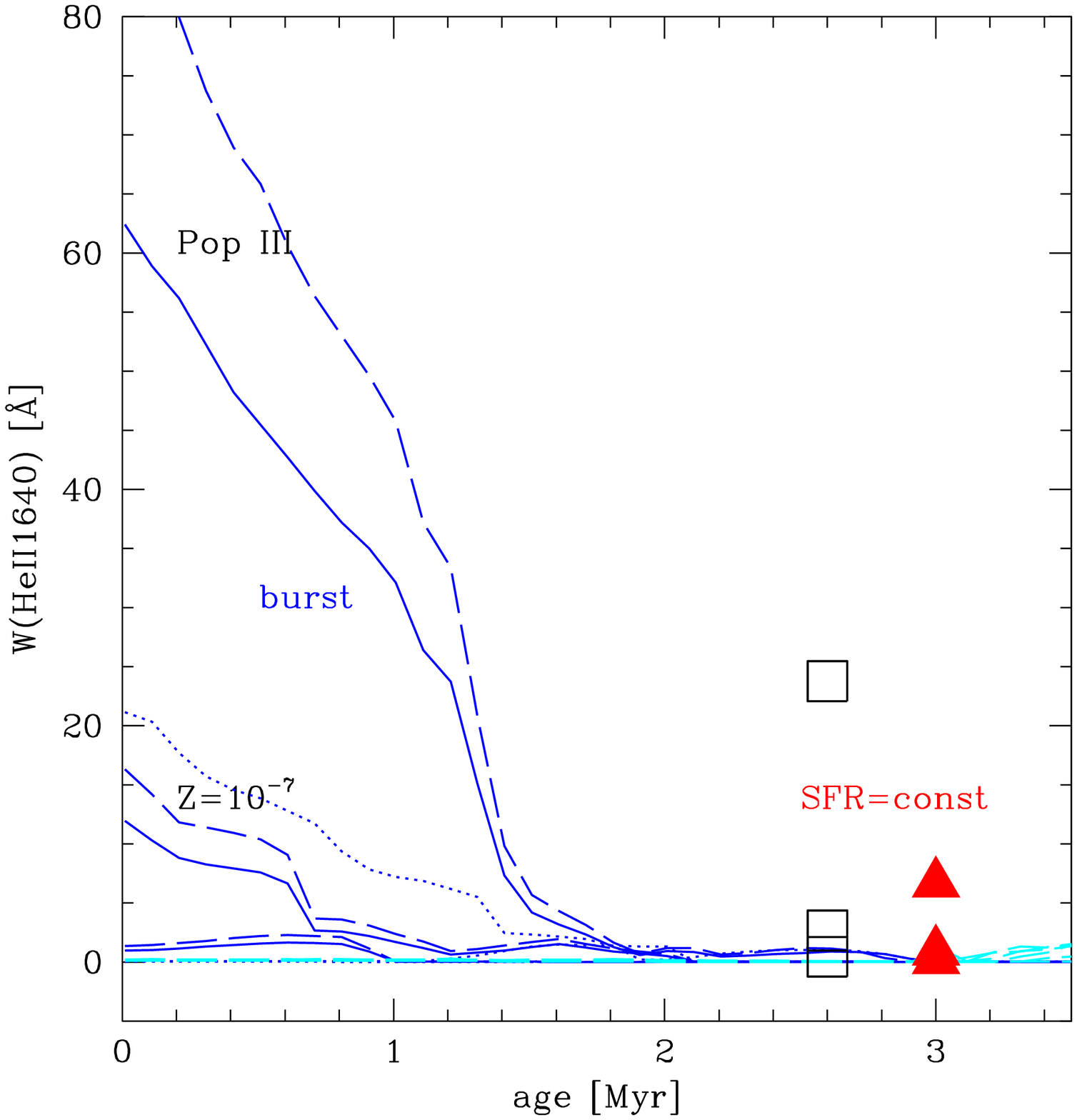,width=8.8cm}}
\caption{Temporal evolution of the \lya\ equivalent width
(left panel) and \Heiiuv\ equivalent width (right) for instantaneous bursts at all 
metallicities.
The very metal-poor models ($Z=$ 0, $10^{-7}$, and $10^{-5}$) with the 
IMFs C (50-500 \msun), B  (1-500 \msun) and A (1-100 \msun) are shown
as short-dashed, solid, and dotted lines respectively from top to bottom. 
The remaining metallicities (for IMF A) are shown with dashed lines.
The equilibrium values for SFR$=const$ at metallicities $Z \le
10^{-5}$ are plotted on the right (at arbitrary ages) using open
squares for the IMF C, filled triangles for IMF B, open circles
for IMF A, and using short lines for higher metallicities (with IMF A).  
Note the very large maximum $W(\lya)$ predicted at young ages.
$W($\Heiiuv)$ \protect\ga$ 5 \AA\ are only expected at the lowest metallicities
($Z \protect\la 10^{-7}$), except if hot WR-like stars not included in the 
tracks were formed e.g.\ through important stellar mass loss (Sect.\ \ref{s_discuss}).
See text for further discussion
}
\label{fig_ew}
\end{figure*}

\section{\lya\ and \Heiiuv\ emission at various metallicities}
\label{s_lya}

To first order recombination line luminosities are proportional 
to the ionising photon flux and 
are thus simply expressed as (in units of erg s$^{-1}$) 
\begin{equation}
  L_l(t) \,\, [{\rm erg \, s^{-1}}] = c_l \,\, (1 - f_{\rm esc}) \,\, 
        Q_i(t) \,\,\, [{\rm s^{-1}}],
  \label{eq_lines}
\end{equation}
where $c_l$ are the line emission coefficients given 
in Table \ref{tab_lines},
$f_{\rm esc}$ is the photon escape fraction out of the galaxy or 
observed aperture,
and $Q_i$ is the ionising photon flux (in units of s$^{-1}$)
corresponding to the appropriate recombination line.

However, as well known \lya\ constitutes a particular case due to its very 
large line optical depth, which implies that several effects
(e.g.\ dust absorption, ISM geometry and velocity structure) 
can alter the total \lya\ emission and lead to complex line profiles
(cf.\ Charlot \& Fall 1993, Valls-Gabaud 1993, 
Chen \& Neufeld 1994, Kunth \etal\ 1998, 
Loeb \& Rybicki 1999, Tenorio-Tagle \etal\ 1999).
Furthermore for \lya\ source at redshifts close to or above the redshift of 
reionisation
the intrinsic \lya\ emission may be further reduced or suppressed 
by absorption in nearby line of sight HI clouds 
(cf.\ Miralda-Escude \& Rees 1997, but also Haiman 2002, Madau 2002).
A proper treatment of these effects requires a complex solution of 
radiation transfer which depends strongly on geometrical properties of 
the ISM and IGM, and for which no general solution is possible.
One must thus caution that depending on the application our
simplifying assumptions may not apply and the predicted \lya\ emission 
should thus be considered as an upper limit. 
Note that these difficulties do not affect other recombination lines
such as \ha\ and \Heiiuv, whose optical depth is strongly reduced compared
to \lya.

Bearing the above in mind, the time evolution of the \lya\ and
\Heiiuv\ line luminosities can be deduced from the evolution of 
\Qh\ and \Qhep\ respectively given in Fig.\ \ref{fig_qi}.

In the case of constant star formation, at equilibrium,
recombination line luminosities $L_l$ are proportional to the star 
formation rate (SFR), i.e.
\begin{equation}
  L_l = (1-f_{\rm esc}) f_l \left(\frac{\rm SFR}{\msun {\rm yr}^{-1}}\right).
\label{eq_sfr}
\end{equation}
For $L_l$ in erg s$^{-1}$ the proportionality constants $f_l$ 
for \lya, \ha, and \Heiiuv\ are listed in columns 9--11 of Table \ref{tab_csf}.
The variations of the ionising flux \Qh\ already discussed above imply in particular
lower star formation rates at low metallicity when identical recombination
lines are used. 
We note that our \ha\ SFR conversion factors are in good agreement with
other computations at various metallicities (e.g.\ Kennicutt 1998, Sullivan 
\etal\ 2001) when rescaled for their \mlow\ using $c_M$ (Table \ref{tab_models}).

The predicted \lya\ and \Heiiuv\ emission line equivalent widths of 
ageing bursts of different metallicities and all the IMF cases
considered are shown in Fig.\ \ref{fig_ew}.
Note that our revised Pop III models show smaller \lya\ equivalent widths
compared to the previous calculation in S02.
This is due to an erroneous continuum definition in the earlier
computations. The new models supersede those of S02.
Good agreement is also found with the $W(\lya)$ predictions of 
Tumlinson \etal\ (2002).
The reader is also reminded that none of these recent calculations 
include stellar \lya\ absorption (cf.\ Valls-Gabaud 1993, Charlot \&
Fall 1993, and Valls-Gabaud \& Schaerer 2002 for new predictions).

Maximum \lya\ equivalent widths of $\sim$ 240--350 \AA\ are predicted 
for metallicities between solar and $4. \, 10^{-4}$. 
For $Z$ down to zero (Pop III), max$(W(\lya))$ may reach up to $\sim$
800--1500 \AA\ for the various adopted IMFs (cf.\ S02)
For comparison, the equilibrium values for SFR=$const$ are in the
range of $W(\lya) \sim$ 175--275, 240--350, 500--930 \AA\ for 
IMF A, B, and C respectively at $Z \le 10^{-5}$,
and $\sim$ 70--100 \AA\ for higher metallicities (IMF A).  
Note that the increased Lyman continuum output of young very metal-poor populations
alone does not explain the strong increase of $W(\lya)$ (cf.\ Table \ref{tab_zams}).
In addition the reduced stellar UV continuous luminosity at $\lambda \sim$ 1215 \AA, 
due to the shift of the SED peak far into the Lyman continuum (Fig.\ \ref{fig_seds}),
also contributes to increase W(\lya).

A high median \lya\ equivalent width ($\sim$ 430 \AA) was found 
in the Large Area Lyman Alpha (LALA) survey of Malhotra \& Rhoads
(2002) at $z=4.5$ and interpreted as due to AGN, starbursts with 
flat IMFs, or even Pop III objects. 
Indeed, if constant star formation is appropriate for their objects
and the IMF slope is universally that of Salpeter, the observed
large $W(\lya)$ would require very metal-poor populations with 
large upper mass cut-offs and/or an increased lower cut-off (e.g.\ 
IMFs B or C).
Alternatively their observations could also be explained by
predominantly young bursts, with metallicities $\la 10^{-5}$
and no need for extreme IMFs (Fig.\ \ref{fig_ew}).
%
This issue will be addressed in detail in a subsequent publication 
(Valls-Gabaud \& Schaerer 2002).

As expected from the softening of the radiation field with increasing
metallicity, the \Heiiuv\ equivalent widths decreases strongly with $Z$;
values $W($\Heiiuv$) \ga$ 5 \AA\ are expected only at metallicities
$Z \la 10^{-7}$,
except if hot WR-like stars (cf.\ Sect.\ \ref{s_discuss})
or non-stellar sources (e.g.\ X-rays, AGN)
provide sufficient amounts of He$^+$ ionising photons.

Part of the relative weakness of  $W($\Heiiuv$)$ compared to $W(\lya)$
is due to a relatively strong, mostly nebular, continuum flux at 1640 \AA\ 
(see S02). As \lya\ emission may be strongly reduced due to the effects
discussed earlier in objects beyond the re-ionisation redshift,
and the \Heiiuv\ luminosity is potentially strong enough to be detected 
(cf.\ Tumlinson \etal\ 2001, Oh \etal\ 2001, Schaerer \& Pell\'o 2001), 
it is a priori not clear if both lines may be observed simultaneously 
and if so which of the two lines would be stronger.

\section{Conclusions}
\label{s_conclude}
We have examined the spectral properties of the ionising continua,
the Lyman-break, and the \lya\ and \Heiiuv\ recombination lines 
in starbursts with metallicities $Z$ from zero -- corresponding to
primordial, Pop III objects -- over low metallicities ($Z \ga 4.\ 10^{-4}$)
observed in nearby galaxies, up to solar metallicity (\zsun$=$0.02).

Our computations, including new sets of stellar evolution 
models at very low metallicities ($Z = 10^{-7}$, $10^{-5}$) and previously
published grids at other $Z$, allow us in particular to study how
spectral properties vary in the transition from Pop III objects to
``normal'' currently measurable metallicities.
 
Various IMFs are treated, including also cases where very massive stars
(up to $\sim 500$ \msun) are formed, as suggested by hydrodynamical calculations
for metallicities $Z \la Z_{\rm crit} \approx 10^{-5}$ (Bromm \etal\ 2001a,
Abel \etal\ 1998, Nakamura \& Umemura 2001).
 
Predictions are provided for the number of H, He$^0$, and He$^+$ 
ionising photons and average photon energies in these continua, 
the hardness of the ionising spectrum, 
the amplitude of the Lyman-break, 
the number of photons able to photodissociate H$_2$,
and finally recombination line luminosities and equivalent widths
(mostly for \lya\ and \Heiiuv).

Two limiting cases of star formation histories, instantaneous bursts and 
constant star formation, are considered.
For SFR=$const$, presumably appropriate to describe the average properties
of starbursts galaxies or populations thereof, the following
main results have been obtained:

\begin{itemize}
\item[$\bullet$] As expected from numerous earlier computations the Lyman 
continuum flux \Qh\ increases with decreasing metallicity.
For a universal Salpeter IMF from 1--100 \msun\ the enhancement
reaches typically a factor of $\sim$ 3 between solar metallicity and Pop III
objects for a constant star formation (Sect.\ \ref{s_qi}, cf.\ Tumlinson
\& Shull 2000, Schaerer 2002).

\item[$\bullet$] While for metallicities $Z \ga 1/50 \zsun$ the amplitude of 
the Lyman-break is rather metallicity independent, 
a reduction by a factor $\sim$ 2 is found at lower metallicities,
related to the strong increase of the average stellar temperature 
(Sect.\ \ref{s_lyb}).

\item[$\bullet$] The predicted hardness of the ionising fluxes between
$\ge 54$ eV and $\ge 13.6$ eV, i.e.\ the ability to doubly ionise He,
decreases by $\sim$ 1.5--2 or more
orders of magnitude from $Z=0$ to $\sim 10^{-4}$ depending strongly
on the upper mass cut-off of the IMF (Sect.\ \ref{s_q2}).

From empirical constraints we derive a hardness $\log \Qrathep \sim$
--3.2 to --2.6 for metal-poor starbursts ($1/25 \la Z/\zsun \la 1/4$)
and softer spectra for higher metallicities (Sect.\ \ref{s_emp}).
The former should provide the best estimate of \Qrathep;
the latter finding is also compatible with recent evolutionary
synthesis models of Smith \etal\ (2002) including line blanketed 
non-LTE atmospheres for WR and O stars.

We also provide (Sect.\ \ref{s_discuss}) a simple estimate of the 
possible impact of hot WR like stars on \Qrathep\ at very low 
metallicities ($Z \la  10^{-4}$). 
Such stars could eventually form due to a strong enhancement of 
mass loss related to rapid rotation (Marigo \etal\ 2002)
or in principle also due very efficient rotational mixing processes
(cf.\ Meynet \& Maeder 2002), although 
we believe that these scenarios are quite unlikely or overall of minor 
importance.

\item[$\bullet$] Finally, calibrations for star formation rate
determinations from \lya\ and other recombination lines at all
metallicities 
and for various IMFs are provided (Sect.\ \ref{s_lya}).

\end{itemize}

For young bursts, the maximum \lya\ equivalent width predicted 
is shown to increase strongly with decreasing metallicity from
$W(\lya) \sim$ 240--350 \AA\ at $Z \ga$ 1/50 \zsun\ to 400--850 \AA\
or higher at $Z$ between $10^{-5}$ and 0 (Pop III) for the same
Salpeter IMF (Sect.\ \ref{s_lya}).

We find that non-negligible \Heiiuv\ emission due to photoionisation
from stellar sources appears to be limited to very small metallicities 
($\log(Z/\zsun) \la -5.3$) and Population III objects, except
if hot WR like stars, whose existence appears very speculative,
were frequent.

The detailed model predictions presented here are available on 
the Web through the CDS and at {\tt http://webast.ast.obs-mip.fr/sfr/}.
In subsequent publications our models are applied to 
the interpretation of \lya\ observations in high redshift galaxies
(Valls-Gabaud \& Schaerer 2002),
modeling of the combined chemical enrichment and re-ionisation history
of the Universe (Ferrara \& Schaerer 2002),
and feasibility studies on the detection of Population III objects
(Pell\'o \& Schaerer 2002).


\acknowledgements{I thank 
Andrea Ferrara, Roser Pell\'o, David Valls-Gabaud for 
stimulating discussions and comments on an earlier version of the 
manuscript. 
Useful comments from Tom Abel, Daniel Kunth, and Grazyna Stasi\'nska 
on the draft and related issues were also appreciated.
Richard Norris kindly provided model results from his calculations for
comparison.
Last, but not least, I thank the referee for useful comments which 
helped to improve the presentation.
}

\end{document}